\documentclass[%
 reprint,
 amsmath,amssymb,
 aps,
]{revtex4-2}

\usepackage{soul}
\usepackage{graphicx}% Include figure files
\usepackage{dcolumn}% Align table columns on decimal point
\usepackage{bm}% bold math
\usepackage[utf8]{inputenc}
\usepackage[T1]{fontenc}
\usepackage{amsthm,amsfonts,mathrsfs}
\usepackage{mathtools}
\usepackage{slashed}
\usepackage{tikz}
\usepackage[compat=1.1.0]{tikz-feynman}
\usepackage[colorlinks=true,linkcolor=blue,citecolor=blue,urlcolor=blue]{hyperref}

\graphicspath{{figures/}}
\newcommand{\feynscale}{0.85}

\newcommand{\ee}{\mathrm{e}}
\newcommand{\ii}{\mathrm{i}}
\newcommand{\dd}{\mathrm{d}}
\newcommand{\Fcal}{\mathcal{F}}
\newcommand{\Lcal}{\mathcal{L}}
\newcommand{\Gcal}{G}
\newcommand{\Ical}{\mathcal{I}}
\newcommand{\Tvec}{\vec{\mathcal{T}}}

\begin{document}

\preprint{APS/123-QED}

\title{Spinor tunnelling amplitudes from all-order Feynman diagram summation}

\author{M. Fleming}
\email{max.fleming@anu.edu.au}
\affiliation{%
 Department of Fundamental and Theoretical Physics,\\
 Australian National University, Canberra, ACT 2601, Australia
}%
\affiliation{%
ARC Centre of Excellence for Dark Matter Particle Physics, Australia}
\author{C. Simenel}
\email{cedric.simenel@anu.edu.au}
\affiliation{%
 Department of Fundamental and Theoretical Physics,\\
 Australian National University, Canberra, ACT 2601, Australia
}%
\affiliation{%
ARC Centre of Excellence for Dark Matter Particle Physics, Australia}
\date{\today}

\begin{abstract}

This work presents a non-perturbative quantum field theoretic computation of spinor tunnelling through a  potential barrier. Within position space potential formalism, we identify coupled recursion relations forming a self-consistent integral equation. Our method, enabling both analytical and numerical approaches, transforms into a formalism admitting both numerical framework for arbitrary potentials and analytical exact results for select barriers. For the rectangular barrier, the QFT-derived amplitudes reproduce the relativistic quantum mechanical (RQM) results, thus establishing a foundation for future calculation of beyond-RQM field-theoretic corrections.

\vspace{1em}
\end{abstract}

\maketitle

%=============================================================
\section{Introduction}
\label{sec:intro}
%=============================================================
 
% Quantum tunnelling is inherently non-perturbative as the below-barrier
% transmission amplitude is exponentially suppressed in the coupling and 
% cannot be reached by any finite truncation of a perturbative series~\cite{dunne_perturbative--nonperturbative_2002,suslov_divergent_2005}.
% This obstruction has prevented a consistent quantum field theoretic treatment
% of single-particle tunnelling, despite the phenomenon's prominence in alpha
% decay~\cite{gamow_zur_1928,geiger_lvii_1911}, stellar nucleosynthesis~\cite{trixler_quantum_2013},
% scanning tunnelling microscopy~\cite{binnig_scanning_1987,tersoff_theory_1985,gottlieb_bardeens_2006} and biological electron
% transfer~\cite{devault_quantum_1980,xin_quantum_2019}.
% Quantum tunnelling is inherently non-perturbative: the below-barrier transmission amplitude is non-analytic in the coupling, vanishing faster than any power, and so lies beyond the reach of any finite order of perturbation theory
% {\color{blue}
Quantum tunnelling is inherently non-perturbative: the below-barrier transmission amplitude has the characteristic form $\exp(-C/g^\alpha)$ (where $C>0$ is determined by the barrier, $g$ is the coupling or interaction strength, and $\alpha$ is a positive constant that depends on the problem), which is non-analytic at zero coupling and therefore cannot be reproduced by any finite-order perturbative expansion~\cite{dunne_perturbative--nonperturbative_2002,suslov_divergent_2005}.
% }
This obstruction has prevented a consistent quantum field theoretic treatment of single-particle tunnelling, despite the phenomenon's prominence in alpha decay~\cite{gamow_zur_1928,geiger_lvii_1911}, stellar nucleosynthesis~\cite{diehl2026}, %\cite{trixler_quantum_2013}, 
scanning tunnelling microscopy~\cite{binnig_scanning_1987,tersoff_theory_1985,gottlieb_bardeens_2006} and biological electron transfer~\cite{devault_quantum_1980,xin_quantum_2019}.

% The motivation for a QFT-native description is substantial.
% The Klein paradox~\cite{klein_reflexion_1929,holstein_kleins_1998} showcases pathologies of single-particle relativistic quantum mechanics (RQM) theories, such as the Dirac equation, which produce anomalous scattering coefficients for step potentials above the Klein threshold $V > 2mc^2$~\cite{villalba_transmission_2003,dombey_seventy_1999}.
% It has long been proposed that this pathology is resolved only once pair production~\cite{hansen_kleins_1981,dombey_seventy_1999}
% is accounted for, 
% which necessitates an interacting field
% theory~\cite{gardiner_tunnelling_2013}.
% Moreover, precision radiative corrections, such as the anomalous magnetic
% moment~\cite{schwinger_quantum-electrodynamics_1948}, the Lamb
% shift~\cite{lamb_fine_1947,bethe_electromagnetic_1947,french_electromagnetic_1949,jentschura_double-logarithmic_2002},
% and vacuum
% polarisation~\cite{uehling_polarization_1935,schwinger_gauge_1951}, which have
% no counterpart in RQM are expected to modify tunnelling
% rates~\cite{flambaum_radiation_1999}.
% However, deploying the full QFT apparatus to capture the true phenomenology
% of tunnelling requires the recalcitrantly abstruse divergence barrier to be
% surmounted~\cite{dyson_divergence_1952,t_hooft_regularization_1972}.

The motivation for a QFT-native description is substantial. The Klein paradox~\cite{klein_reflexion_1929,holstein_kleins_1998} showcases pathologies of single-particle relativistic quantum mechanics (RQM) theories, such as the Dirac equation, which predict paradoxical transmission through step potentials exceeding the Klein threshold $V > 2mc^2$, where the reflection coefficient can exceed unity~\cite{villalba_transmission_2003,dombey_seventy_1999}.
It has long been proposed that this pathology is resolved only once pair production~\cite{hansen_kleins_1981,dombey_seventy_1999} is accounted for, which necessitates an interacting field theory~\cite{gardiner_tunnelling_2013}.
Moreover, precision radiative corrections, such as the anomalous magnetic moment~\cite{schwinger_quantum-electrodynamics_1948}, the Lamb shift~\cite{lamb_fine_1947,bethe_electromagnetic_1947,french_electromagnetic_1949,jentschura_double-logarithmic_2002}, and vacuum polarisation~\cite{uehling_polarization_1935,schwinger_gauge_1951}, which have no counterpart in RQM, are expected to modify tunnelling rates~\cite{flambaum_radiation_1999}.
However, deploying the full QFT apparatus requires confronting the divergences that afflict the perturbative expansion~\cite{dyson_divergence_1952,t_hooft_regularization_1972}.

% For scalar fields, an all-order resummation was achieved by Zieli\'{n}ski
% \textit{et al.}~\cite{zielinski_tunnelling_2024} for potentials composed of
% delta functions, where the perturbative series forms a geometric progression.
% The scalar framework was further developed in
% Ref.~\cite{zielinski_quantum_2024} to include loop corrections from a
% quantised mediator field.
% Spinors resist this approach on two counts: delta-function potentials are
% pathological for spin-$\tfrac{1}{2}$
% particles~\cite{subramanian_relativistic_1971,calkin_proper_1987,coutinho_unusual_2009},
% and the matrix-valued numerator $\slashed{p}+m$ of the Dirac propagator
% prevents a direct scalar-like resummation.
% Perturbative QFT  calculations for spinors in rectangular barriers are available
% to second order~\cite{de_leo_potential_2009,xu_path_2016}, but no all-order
% result has existed prior to this work.

For scalar fields, an all-order resummation was achieved by Zielinski \textit{et al.}~\cite{zielinski_tunnelling_2024} for potentials composed of delta functions, where the perturbative series forms a geometric progression. The scalar framework was further developed in Ref.~\cite{zielinski_quantum_2024} to include loop corrections from a quantised mediator field.
Spinors resist this approach on two counts: delta-function potentials are pathological for spin-$\tfrac{1}{2}$ particles~\cite{subramanian_relativistic_1971,calkin_proper_1987,coutinho_unusual_2009} and the matrix-valued numerator $\slashed{p}+m$ of the Dirac propagator prevents a direct scalar-like resummation.
Perturbative QFT tunnelling calculations for spinors in rectangular barriers are available to second order~\cite{de_leo_potential_2009,xu_path_2016}, but no all-order result has existed prior to this work.

% Existing theoretical approaches to tunnelling beyond single-particle quantum
% mechanics fall into two broad classes.
% In field theory, semiclassical instanton methods describe tunnelling between
% degenerate vacua or the decay of metastable
% states~\cite{coleman_fate_1977,callan_fate_1977}; these describe 
% tunnelling of a complete field configuration rather than the
% phenomenologically distinct particle tunnelling 
% transmissions through spatial potential barriers.
% Resurgent transseries techniques have further developed this
% direction~\cite{basar_resurgence_2013,dunne_resurgence_2012,dorigoni_introduction_2019}.
% For actual particle tunnelling in many-body systems: coupled-channels 
% and time-dependent Hartree--Fock (TDHF) approaches
% are the workhorses for sub-barrier nuclear
% fusion
% and fission~\cite{balantekin_quantum_1998,simenel_nuclear_2012,bonasera_calculation_2020},
% while imaginary-time mean-field
% extensions~\cite{mcglynn_imaginary-time_2020} and
% Caldeira--Leggett-type dissipative
% models~\cite{caldeira_quantum_1983,hanggi_reaction-rate_1990}
% address collective tunnelling and environment-coupled systems, respectively.
% Such frameworks, however, originate from within fixed-particle-number sectors
% and cannot systematically incorporate pair creation, vacuum polarisation or
% radiative correction effects that require a fully interacting quantum field
% theory.

Existing theoretical approaches to tunnelling beyond single-particle quantum mechanics fall into two broad classes.
In field theory, semiclassical instanton methods describe tunnelling between degenerate vacua or the decay of metastable states~\cite{coleman_fate_1977,callan_fate_1977}; these describe tunnelling of a complete field configuration rather than the transmission of a single particle through a spatial barrier.
Resurgent transseries techniques have further developed this direction~\cite{basar_resurgence_2013,dunne_resurgence_2012,dorigoni_introduction_2019}.
% {\color{blue}
Sub-barrier nuclear fusion is often described with the coupled-channel model~\cite{balantekin_quantum_1998,simenel2013b} or from energy-dependent microscopic potentials \cite{umar2014a,simenel_nuclear_2025}, 
% }
%For genuine single-particle tunnelling in many-body systems, coupled-channels and time-dependent Hartree--Fock (TDHF) approaches are the workhorses for sub-barrier nuclear fusion and fission~\cite{balantekin_quantum_1998,simenel_nuclear_2012,bonasera_calculation_2020}, 
while imaginary-time mean-field approaches~\cite{levit1980c,mcglynn_imaginary-time_2020} and Caldeira--Leggett-type dissipative models~\cite{caldeira_quantum_1983,hanggi_reaction-rate_1990} address collective tunnelling and environment-coupled systems, respectively.
Such frameworks, however, operate within fixed-particle-number sectors and cannot systematically incorporate pair creation, vacuum polarisation or radiative correction effects that require a fully interacting quantum field theory.

This work shows that by working directly with position-space scattering
densities 
% {\color{blue}
(whose spatial integrals give the scattering amplitudes), it is possible to establish a non-perturbative 
quantum field theoretic summation technique.
The key observation is that the $n$-th order density satisfies a recursion
relation expressed in terms of $(n{-}1)$-th order contributions.
Summing the resulting Neumann series yields a Schwinger--Dyson-type integral
equation~\cite{curtis_truncating_1990,dyson_divergence_1952,bashir_dynamical_2012}
that, upon differentiation, reduces to a system of two coupled first-order
ODEs amenable to standard numerical methods and, for the rectangular barrier,
to exact analytic solution.
The QFT amplitudes reproduce the RQM
results~\cite{de_leo_above_2006,cotaescu_applying_2007},
thus providing a framework for introducing field-theoretic corrections to
descriptions of tunnelling in many-particle interacting
systems~\cite{simenel_heavy-ion_2018}.

% {\color{blue}
The perturbative expansion is introduced in Sec.~\ref{sec:perturbative}, while the non-perturbative resummation is the subject of Sec.~\ref{sec:resummation}. Applications to a square barrier are presented in Sec.~\ref{sec:results} before concluding in Sec.~\ref{sec:conclusions}.
% }

%=============================================================
\section{Perturbative expansion and Feynman rules}
\label{sec:perturbative}
%=============================================================
 
We consider a Dirac spinor $\psi$ of mass $m$ coupled to a static external
electrostatic potential $V(z)$ depending only on the longitudinal coordinate
$z \equiv x^3$ (superscript numerals
denote Lorentz indices throughout). 
The Lagrangian density is
\begin{equation}
  \Lcal = \bar\psi(\ii\slashed\partial - m)\psi
        - \mu V(z)\,\bar\psi\gamma^0\psi\,.
  \label{eq:lagrangian}
\end{equation}
The scattering amplitude of a particle from initial state $|p,s\rangle$ to final state $|k,s'\rangle$ is given by the S-matrix element $ \langle k,s' | S | p,s \rangle$, where $p=(E,0,0,p^3)$ and $k$ are on-shell momenta, and $\{s,s'\}$ denote spins. 
The electrostatic potential does not affect spin, leading to  $ \langle k,s' | S | p,s \rangle\propto\delta_{ss'}$, so the spin will be usually omitted from the notation. 

Diagrammatically, the S-matrix element is represented as
\begin{equation}
  \langle k | S | p \rangle
  \;=\;
  \begin{tikzpicture}[baseline={([yshift=-.5ex]current bounding box.center)},
                      scale=0.9, transform shape]
    \begin{feynman}
      \vertex (p_in);
      \vertex [right=1.0cm of p_in] (s_in);
      \vertex [right=1.0cm of s_in] (s_out);
      \vertex [right=1.0cm of s_out] (k_out);
      \diagram*{
        (p_in) -- [fermion, edge label'=$p$] (s_in),
        (s_out) -- [fermion, edge label'=$k$] (k_out)
      };
      \coordinate (mid) at ($(s_in)!0.5!(s_out)$);
      \node[draw, circle, fill=teal!30, minimum size=1.0cm] at (mid) {S};
    \end{feynman}
  \end{tikzpicture}\;,
\end{equation}
where the central blob contains all interactions with the external field.
The transmission amplitude  is given as
\begin{equation}
  T = \iint_{-\infty}^{+\infty} \frac{\dd k^1 \dd k^2}{(2\pi)^2}\int_0^{+\infty}\!\frac{\dd k^3}{2\pi}\frac{1}{2E}\,
      \langle k|S|p\rangle\,.
%  R = \int_{-\infty}^0\!\frac{\dd k^3}{2\pi\cdot 2k^0}\,
%      \langle k|S|p\rangle\,.
  \label{eq:TR}
\end{equation}
Similarly, the reflection amplitude $R$ is obtained by integrating the longitudinal outgoing  momentum $k^3$ from $-\infty$ to 0.
%where energy conservation combined with the on-shell condition restricts $k^3 = \pm p^3$.

%Here $u_s(p)$ and $\bar{u}_{s'}(k)$ are the Dirac spinors for the incoming and outgoing
%on-shell particles, satisfying $(\slashed{p} - m) u_s(p) = 0$ and
%$\bar{u}_{s'}(k) (\slashed{k} - m) = 0$. 

In the interaction picture the S-matrix admits the Dyson expansion~\cite{dyson_radiation_1949}
\begin{equation}
  \langle k|S|p\rangle = \sum_{n=0}^\infty \langle k|S|p\rangle^{(n)}\,,
\end{equation}
where $\langle k|S|p\rangle^{(n)}$ involves exactly $n$ insertions of
$\mathcal{H}_{\rm int} = \mu\bar\psi\gamma^0 V(z)\psi$.

The Feynman rules for a spinor coupled to the classical field
$\mathcal{A}_\mu=(V(z),\mathbf{0})$ are:
\begin{enumerate}
  \item \textbf{External-field vertex}:
    \begin{equation}
      \begin{tikzpicture}[baseline={([yshift=-.5ex]current bounding box.center)},
                          scale=0.8, transform shape]
        \begin{feynman}
          \vertex (v);
          \vertex[above=1.4cm of v, inner sep=0pt, outer sep=0pt] (x)
                 {\Large$\otimes$};
          \vertex[left=1.4cm of v] (a);
          \vertex[right=1.4cm of v] (b);
          \diagram*{
            (a)--[fermion, edge label=$p$](v)
                --[fermion, edge label=$k$](b),
            (v)--[photon](x)
          };
        \end{feynman}
      \end{tikzpicture}
      \;=\; -\ii\mu\,\gamma^0\,\tilde V(k-p)\,,
      \label{eq:vertex}
    \end{equation}
    where $\tilde V(q)$ is the Fourier transform of the potential.
 
  \item \textbf{Internal propagator}:
    $G_F(q) = \ii(\slashed{q}+m)/(q^2-m^2+\ii\epsilon)$ is assigned to each
    internal fermion line.
 
  \item \textbf{External spinors}: Incoming particles contribute the Dirac spinor $u_s(p)$;
    outgoing particles contribute $\bar{u}_{s'}(k)$.
 
  \item \textbf{Internal momenta} $q$ are integrated over with $\int\!\dd^4 q/(2\pi)^4$ per 
    internal line.
 
  \item \textbf{Momentum conservation}: Each vertex conserves four-momentum,
    but the external classical potential~\eqref{eq:vertex} can supply
    arbitrary three-momentum.
    (Time-translation invariance of the static potential enforces energy
    conservation at every vertex.)
\end{enumerate}

%The zeroth- and first-order contributions are
%\begin{align}
%  \langle k|S|p\rangle^{(0)}
%    &= (2\pi)^4\,\delta^{(4)}(k-p)\,\delta_{ss'}\,,
%  \label{eq:S0}\\
%  \langle k|S|p\rangle^{(1)}
%    &= -\ii\mu\,\bar{u}_{s'}(k)\,\gamma^0\,u_s(p)\;\tilde V(k-p)\,,
%  \label{eq:S1}
%\end{align}
%written with spin indices, which hereafter are assumed implicit.
The general $n$-th order term, with $n$ vertices connected by $n{-}1$
internal propagators and convention $q_0\equiv p$ %, $q_n\equiv k$, 
reads

\begin{widetext}
\begin{subequations}
\begin{align}
  \langle k|S|p\rangle^{(n)}
  &=
  \scalebox{\feynscale}{\begin{tikzpicture}[
    baseline={([yshift=0.3ex]current bounding box.center)}]
    \begin{feynman}
      \vertex (e1) at (-5,-1);
      \vertex (p1) at (-4, 0);
      \vertex (p2) at (-2, 0);
      \vertex (p3) at ( 0, 0);
      \vertex (p5) at ( 2, 0);
      \vertex (p6) at ( 4, 0);
      \vertex (e2) at ( 5,-1);
      \vertex[inner sep=0pt,outer sep=0pt] (x1) at (-4,1) {\Large$\otimes$};
      \vertex[inner sep=0pt,outer sep=0pt] (x2) at (-2,1) {\Large$\otimes$};
      \vertex[inner sep=0pt,outer sep=0pt] (x3) at ( 0,1) {\Large$\otimes$};
      \vertex[inner sep=0pt,outer sep=0pt] (x5) at ( 2,1) {\Large$\otimes$};
      \vertex[inner sep=0pt,outer sep=0pt] (x6) at ( 4,1) {\Large$\otimes$};
      \diagram*{
        (e1)--[fermion,momentum'=$p\equiv q_0$](p1)
            --[fermion,momentum'=$q_1$](p2)
            --[fermion,momentum'=$q_2$](p3)
            --[draw=none,edge label=$\boldsymbol{\cdots}$](p5)
            --[fermion,momentum'=$q_{n-1}$](p6)
            --[fermion,momentum'=$k$](e2),
        (p1)--[photon](x1),(p2)--[photon](x2),(p3)--[photon](x3),
        (p5)--[photon](x5),(p6)--[photon](x6),
      };
    \end{feynman}
  \end{tikzpicture}}
  \label{eq:nth_diagram} \\
  &= (-\ii\mu)^n
    \int\!\frac{\dd^4 q_1}{(2\pi)^4}\cdots\frac{\dd^4q_{n-1}}{(2\pi)^4}\,\,
    \bar u(k)\,\gamma^0\,\tilde V(k-q_{n-1})\,\,
    \mathcal{P}\left[\prod_{j=1}^{n-1}G_F(q_j)\,\gamma^0\,
    \tilde V(q_j-q_{j-1})\right]u(p)\,,
  \label{eq:nth_amplitude}
  \nonumber\\
%   &= \int \dd z_1\cdots\dd z_n\, (-\ii\mu)^n
%     \int\!\frac{\dd^4 q_1}{(2\pi)^4}\cdots\frac{\dd^4q_{n-1}}{(2\pi)^4}\,\,
%     \bar u(k)\,\gamma^0\,\left(\prod_{i=0}^{2}2\pi\delta(k^i-q^i_{n-1})\right)\;
%    V(z_{n})\,\ee^{-\ii (k^3-q_{n-1}^3) z_{n}}\nonumber\\
% &\times    \mathcal{P}\left[\prod_{j=1}^{n-1}G_F(q_j)\,\gamma^0\,
%     \left(\prod_{i=0}^{2}2\pi\delta(q_j^i-q^i_{j-1})\right)\;
%    V(z_j)\,\ee^{-\ii (q_j^3-q_{j-1}^3) z_{j}}\right]u(p)\,,
%   \label{eq:nth_amplitude_expanded}
\end{align}
\end{subequations}
\end{widetext}
where $\mathcal{P}$ denotes the path ordering operator so that $j=1$ stands
rightmost.

For the static one-dimensional potential, the Fourier transform factorises as
% , and 
\begin{equation}
  \tilde V(q) = \left(\prod_{i=0}^{2}2\pi\delta(q^i)\right)\;
    \int\!\dd z\;V(z)\,\ee^{-\ii q^3 z}\,.
  \label{eq:Vtilde}
\end{equation}
% was used in the last line.

Inserting Eq.~\eqref{eq:Vtilde} at every vertex of Eq.~\eqref{eq:nth_amplitude} renders the $n$-th order amplitude an integral over the $n$ vertex positions,
\begin{equation}
 \langle k|S|p\rangle^{(n)}=\int \dd z_1\cdots \dd z_n\,\, \mathcal{S}_{kp}^{(n)}(\vec{z})\,,
 \label{eq:density_def}
\end{equation}
which defines the $n$-th order scattering density $\mathcal{S}_{kp}^{(n)}(\vec{z})$, whose explicit form is given in Eq.~\eqref{eq:density_explicit} of Appendix~\ref{app:densities}.
The densities, rather than the amplitudes themselves, are the subjects of resummation in Sec.~\ref{sec:resummation}.

Although the diagrammatic structure~\eqref{eq:nth_diagram} is formally identical to the scalar ladder~\cite{zielinski_tunnelling_2024}, the integrals in~\eqref{eq:nth_amplitude} do not collapse into a geometric series. The matrix-valued propagator numerators $\slashed{q}_j + m$ do not commute through the $\gamma^0$ vertices, so the standard scalar resummation $\sum_n x^n = (1-x)^{-1}$ has no direct analogue.
The route taken in the next section circumvents this obstruction by first integrating out the longitudinal loop momentum at each internal line, reducing the matrix structure to the closed-form kernel~\eqref{eq:G_result} before the remaining $z_j$ integrals are addressed.

%=============================================================
\section{Non-perturbative resummation}
\label{sec:resummation}
%=============================================================

\subsection{Coupled recursion relation for the densities}

The $\delta-$functions of
Eq.~\eqref{eq:Vtilde} are substituted into Eq.~\eqref{eq:nth_amplitude} at each vertex, enforcing energy and transverse momentum conservation. As a result, particles transmitted through (reflected by) the barrier have a momentum $p_+\equiv(\hat{p},p^3)$
($p_-\equiv(\hat{p},-p^3)$), where $\hat{p}=(p^0,p^1,p^2)$. 

%\subsection{One-dimensional propagator structure}
%\label{sec:1d_prop}
%For the one-dimensional static potential, the Fourier transform factorises as
%\begin{equation}
%  \tilde V(q) = (2\pi)^3\prod_{i=0}^{2}\delta(q^i)\;
%    \int\!\dd z\;V(z)\,\ee^{-\ii q^3 z}\,.
%  \label{eq:Vtilde}
%\end{equation}
%The three delta functions at each vertex in Eq.~\eqref{eq:nth_amplitude} enforce energy and transverse
%momentum conservation, fixing every internal momentum to the form
%$q_j = (p^0,0,0,q_j^3)=(\hat{p},q^3_j)$ and constraining the amplitude to vanish unless
%$\hat{k}=\hat{p}$.
%Thanks to the on-shell condition $(p^0)^2=m^2+(p^3)^2$, $ \hat{\Gcal}$ can be expressed as a function of the longitudinal momentum $p^3$:

The only non-trivial internal integrations are those over the longitudinal
momenta $q^3_j$.
They involve the Fourier transform of the Feynman propagator with respect to
its longitudinal momentum,
\begin{equation}
  \hat{\Gcal}(\hat{k},\xi) =
    \int\!\frac{\dd q^3}{2\pi}\;\ii\,
    \frac{\hat{\slashed{k}} - q^3\gamma^3 + m}
         {(k^3)^2-(q^3)^2+\ii\epsilon}\;
    \ee^{\ii q^3\xi}\,,
\label{eq:Ghat_def}
\end{equation}
where $\hat{\slashed{k}}=k^0\gamma^0-k^1\gamma^1-k^2\gamma^2$ and  $(\hat{k})^2=(k^0)^2-(k^1)^2-(k^2)^2=m^2+(k^3)^2$ as the final momentum $k$ is on-shell.

Evaluating~\eqref{eq:Ghat_def} by contour integration, Jordan's lemma is
applied to close the contour in the upper (lower) half-plane for
$\xi>0$ ($\xi<0$) so that the arc contribution vanishes.
This distinguishes two cases of residues at $q^3=\pm k^3$, yielding
\begin{align}
   \hat{\Gcal}(\hat{k},\xi) 
  &= \Theta(\xi)\;\frac{\slashed{k}_++m}{2k^3}\;
    \ee^{\ii k^3\xi}\nonumber\\
  &+ \Theta(-\xi)\;\frac{\slashed{k}_-+m}{2k^3}\;
    \ee^{-\ii k^3\xi}\,,
  \label{eq:G_result}
\end{align}
where $k_\pm\equiv(\hat{k},\pm k^3)$.
%
%$(ii)$ Although the right-hand-side of Eq.~\eqref{eq:G_result} is expressed as a function of the longitudinal momentum $k^3$, $\hat{\Gcal}(\hat{k},\xi)$ does not explicitly depends on $k^3$. This latter remark is important in the next step where we wish to evaluate the term  $ \bar u(k_\pm)\,\Ical(z_n,z_{n-1},k_\pm,q_{n-2})$ in Eq.~\eqref{eq:density_amplitude}. 
%Indeed, while $k^3$ will have to be replaced by $k_\pm^3=\pm k^3$ in Eq.~\eqref{eq:Idensity}, $k^3$ will remain unchanged in Eq.~\eqref{eq:G_result}. 
%This structure is closely related to decompositions used in RQM treatments of piecewise-constant potentials~\cite{cotaescu_applying_2007}.
%The crucial feature of~\eqref{eq:G_result} is that the loop integration over
%$q^3$ has eliminated the matrix-valued denominator of $G_F$, leaving a sum
%of two on-shell projectors weighted by simple plane waves in $\Delta z$.
%This is precisely the structure that will permit a closed-form recursion
%in position space.
%In Eq.~\eqref{eq:Idensity}, $\hat\Gcal$ is expressed as a function of the outgoing momentum $k$ instead of $p$. As mentioned earlier, $\langle k|S|p\rangle$ is non-zero only when $k$ and $p$ are identical up to a sign for the longitudinal component. However, this comes as a consequence of the $\delta$-functions and we thus keep $k$ general, i.e., we do not assume $\hat{k}=\hat{p}$ in the intermediate steps of the calculation. 
%
Rewriting the numerators of~\eqref{eq:G_result} with the spinor completeness relation $\slashed{k}_\pm + m = \sum_s u_s(k_\pm)\,\bar u_s(k_\pm)$ and using the standard inner products
\begin{align}
  \bar u_r(k_\pm)\,\gamma^0\,u_s(k_\pm) &= 2k^0\delta_{rs}\,\,\,\mbox{and}\\
  \bar u_r(k_\pm)\,\gamma^0\,u_s(k_\mp) &= 2m\delta_{rs}\,,
  \label{eq:spinor_ips}
\end{align}
we obtain
\begin{align}
  \bar u(k_\pm)\,\gamma^0\,\Gcal(\hat{k},\xi)
  &= \Theta(\pm\xi)\;\frac{k^0}{k^3}\;
     \ee^{\pm\ii k^3\xi}\,\bar u(k_\pm)
  \nonumber\\
  &\quad+ \Theta(\mp\xi)\;\frac{m}{k^3}\;
     \ee^{\mp\ii k^3\xi}\,\bar u(k_\mp)\,,
  \label{eq:G_result_simp}
\end{align}

Applying Eq.~\eqref{eq:G_result_simp} to the outermost internal line of the
density $\mathcal{S}_{kp}^{(n)}$ expresses it in terms of the $(n{-}1)$-th
order densities.
The calculation, detailed in Appendix~\ref{app:recursion}, yields the
coupled two-channel recursion relation
\begin{widetext}
\begin{equation}
  \begin{bmatrix}
    \mathcal{S}_{k_+ p}^{(n)}(\vec{z})\\[4pt]
    \mathcal{S}_{k_- p}^{(n)}(\vec{z})
  \end{bmatrix}
  = \frac{-\ii\mu V(z_n)}{k^3}
    \begin{bmatrix}
      k^0\,\Theta(z_n{-}z_{n-1}) &
      m\,\ee^{-2\ii k^3 z_n}\,\Theta(z_{n-1}{-}z_n) \\[4pt]
      m\,\ee^{2\ii k^3 z_n}\,\Theta(z_n{-}z_{n-1}) &
      k^0\,\Theta(z_{n-1}{-}z_n)
    \end{bmatrix}
    \begin{bmatrix}
       \mathcal{S}_{k_+ p}^{(n-1)}(\vec{z})\\[4pt]
     \mathcal{S}_{k_- p}^{(n-1)}(\vec{z})
    \end{bmatrix}
%= \Sigma_k(z_k,z_{n-1})        
%\begin{bmatrix}
%       \mathcal{S}_{k_+ p}^{(n-1)}(\vec{z})\\[4pt]
%     \mathcal{S}_{k_- p}^{(n-1)}(\vec{z})
%    \end{bmatrix}
  \label{eq:recursion}
\end{equation}
\end{widetext}

where $\mathcal{S}_{k_\pm p}^{(n)}$ denotes the density evaluated at
outgoing momentum $k_\pm$.

% Let us define the integral operator $\boldsymbol\Sigma$ with density
% \begin{equation}
%   \Sigma(\xi,\eta) =
%   \frac{-\ii\mu\, V(\xi)}{p^3}
%   \begin{bmatrix}
%     p^0\,\Theta(\xi{-}\eta) &
%     m\,\ee^{-2\ii p^3\xi}\,\Theta(\eta{-}\xi)\\[4pt]
%     m\,\ee^{2\ii p^3\xi}\,\Theta(\xi{-}\eta) &
%     p^0\,\Theta(\eta{-}\xi)
%   \end{bmatrix},
%   \label{eq:Sigma}
% \end{equation}
% so that
% $(\boldsymbol\Sigma\cdot\vec{f}\,)(\xi)
%  =\int\!\dd\eta\;\Sigma(\xi,\eta)\,\vec{f}(\eta)$.

\subsection{Transmission and reflection}
The recursion Eq.~\eqref{eq:recursion} relates 
$\mathcal{S}^{(n)}$ to $\mathcal{S}^{(n-1)}$
%the $n$-order density to the order-$n-1$ density, 
through a kernel matrix that depends only on the outermost $z_n$ and $z_{n-1}$ coordinates.
This structure invites a compact formulation. 
%Collecting the transmission and reflection densities,
%with all but the outermost coordinate integrated, into the single-variable
%vector
Defining
\begin{equation}
  \vec{\mathcal{T}}_{kp}^{(n)}(z_n)
  \equiv \int\!\dd z_1\cdots\dd z_{n-1}\,
  \begin{bmatrix}
    \mathcal{S}^{(n)}_{k_+ p}(\vec z)\\[4pt]
    \mathcal{S}^{(n)}_{k_- p}(\vec z)
  \end{bmatrix}\,,%_{z_n=\xi}\,,
  \label{eq:Tvec_def}
\end{equation}
we introduce the integral operator $\boldsymbol\Sigma$ whose kernel is read
off directly from Eq.~\eqref{eq:recursion} with $z_n\to\xi$ and $z_{n-1}\to\eta$,
\begin{equation}
  \Sigma_k(\xi,\eta) =
  \frac{-\ii\mu\, V(\xi)}{k^3}
  \begin{bmatrix}
    k^0\,\Theta(\xi{-}\eta) &
    m\,\ee^{-2\ii k^3\xi}\,\Theta(\eta{-}\xi)\\[4pt]
    m\,\ee^{2\ii k^3\xi}\,\Theta(\xi{-}\eta) &
    k^0\,\Theta(\eta{-}\xi)
  \end{bmatrix},
  \label{eq:Sigma}
\end{equation}
% acting as
% $(\boldsymbol\Sigma\cdot\vec{f}\,)(\xi)
%  =\int\!\dd\eta\;\Sigma(\xi,\eta)\,\vec{f}(\eta)$.
acting as
$(\boldsymbol\Sigma_k\cdot\vec{f}\,)(\xi)
 =\int\!\dd\eta\;\Sigma_k(\xi,\eta)\,\vec{f}(\eta)$.

% The first-order seed vector is
% \begin{equation}
%   \vec{\mathcal{T}}^{(1)}(\xi) =
%   \frac{-\ii\mu\, V(\xi)}{p^3}
%   \begin{bmatrix}p^0\\[2pt] m\,\ee^{2\ii p^3\xi}\end{bmatrix},
%   \label{eq:seed}
% \end{equation}
% and the recursion~\eqref{eq:recursion} gives
% $\vec{\mathcal{T}}^{(n)}=\boldsymbol\Sigma^{n-1}\cdot
%  \vec{\mathcal{T}}^{(1)}$,
% so the total S-matrix vector is the Neumann series
% $\vec{\mathcal{T}}_\infty
%  = \sum_{n=1}^\infty\boldsymbol\Sigma^{n-1}\cdot
%    \vec{\mathcal{T}}^{(1)}$.
%    The sum starts at $n=1$, where the zeroth-order term~\eqref{eq:S0} carries no power of $V$ and therefore does not constitute a scattering contribution, instead furnishing only the unit term in $T=1+\int\!\dd\xi\,T_\infty$.
% Extending~\eqref{eq:Sigma} to act on an $n=0$ seed is in any case obstructed by the step functions undefined argument evaluation and any sensible prescribed seed vectors would still have to contend with the formally ill-defined distributional product $\Theta(\xi-\eta)\,\delta(\xi-\eta)$.

 From Eqs.~\eqref{eq:recursion}, \eqref{eq:Tvec_def} and \eqref{eq:Sigma}, we can write
\begin{equation}
  \vec{\mathcal{T}}_{kp}^{(n)}(\xi)=\int \dd \eta\, \Sigma_k(\xi,\eta) 
\,  \vec{\mathcal{T}}_{kp}^{(n-1)}(\eta),
\end{equation}
% or, equivalently, 
% %Integrating~\eqref{eq:recursion} over the intermediate coordinate
% %$z_{n-1}=\eta$ then collapses the recursion to the single-step relation
% \begin{equation}
% \vec{\mathcal{T}}_{kp}^{(n)}=\boldsymbol\Sigma_k\cdot\vec{\mathcal{T}}_{kp}^{(n-1)}.    
% \end{equation}
% %The iteration terminates on the first-order seed
% %\begin{equation}
% %  \vec{\mathcal{T}}^{(1)}(\xi) =
% %  \frac{-\ii\mu\, V(\xi)}{p^3}
% %  \begin{bmatrix}p^0\\[2pt] m\,\ee^{2\ii p^3\xi}\end{bmatrix},
% %  \label{eq:seed}
% %\end{equation}
% %which is the $n=1$ density evaluated at $z_1=\xi$. 
% Iterating  gives
% \begin{equation}
%     \vec{\mathcal{T}}_{kp}^{(n)}=\boldsymbol\Sigma_k^{n-1}\cdot\vec{\mathcal{T}}_{kp}^{(1)}.\label{eq:TnT1}
% \end{equation}
which may then be iterated down to first order,
\begin{equation}
    \Tvec_{kp}^{(n)}=\boldsymbol\Sigma_k\cdot\Tvec_{kp}^{(n-1)}
    =\boldsymbol\Sigma_k^{\,n-1}\cdot\Tvec_{kp}^{(1)}\,.
    \label{eq:TnT1}
\end{equation}

The iteration terminates on the first-order seed $\Tvec_{kp}^{(1)}$, which
follows from the single-vertex diagram.
Its explicit form, and the integration over the final-state momentum in
Eq.~\eqref{eq:TR}, are carried out in Appendix~\ref{app:seed}.
For an incident particle with $p=(E,0,0,p^3)$ and $p^3>0$, the transmission
and reflection amplitudes~\eqref{eq:TR} are given by summing
Eq.~\eqref{eq:TnT1} over all orders and integrating over the final-state
momentum,
\begin{equation}
\begin{pmatrix} T-1\\ R \end{pmatrix}\!(E)
  = \!\int\!\frac{\dd^3k}{(2\pi)^3}\frac{\Theta(k^3)}{2k^0}\!\int\!\dd \xi
\sum_{n=1}^\infty\left(\boldsymbol\Sigma_{k}^{\,n-1}
\Tvec_{kp}^{(1)}\right)\!(\xi)\,.
  \label{eq:TRassembly_main}
\end{equation}
where the zeroth-order terms of the Dyson
expansion~\eqref{eq:zeroth_order} carry no powers of $V$ and furnish
only the unit contribution to $T$, which we make explicit.
% {\color{blue}
%The $k$ integration is eliminated in Appendix~\ref{app:seed}: the $\delta$
%functions carried by the seed reduce it, order by order, to a pointwise
%evaluation at $k=p_+$, so that
%\begin{equation}
%\begin{pmatrix} T-1\\ R \end{pmatrix}\!(E)
%  = \int\!\dd \xi \; \vec\tau_{\infty}(\xi)\,,
%  \label{eq:TRfromdensity}
%\end{equation}
%where
%\begin{equation}
%   \vec\tau_{\infty}(\xi)
%    \equiv \sum_{n=1}^\infty\left(\boldsymbol\Sigma^{\,n-1}\cdot \vec\tau^{(1)}\right)\!(\xi)
%  \label{eq:neumann}
%\end{equation}
%is the all-order Neumann series, built on the reduced kernel
%$\boldsymbol\Sigma\equiv\boldsymbol\Sigma_{p_+}$ [Eq.~\eqref{eq:Sigma}
%with $k^0\to p^0$, $k^3\to p^3$] and the reduced seed
%\begin{equation}
%  \vec\tau^{\,(1)}(\xi)=
%  \frac{-\ii\mu\, V(\xi)}{p^3}
%  \begin{bmatrix}p^0\\[2pt] m\,\ee^{2\ii p^3\xi}\end{bmatrix}.
%  \label{eq:seed}
%\end{equation}
 %Explicitly, in terms of the iterated kernel
%$(\Sigma_{p_+}^{\,n-1})(\xi,\eta)$, with
%$(\Sigma_{p_+}^{\,0})(\xi,\eta)=\delta(\xi{-}\eta)\,\mathbb{I}$,
Using the expression for $\Tvec_{kp}^{(1)}$ in Eq.~\eqref{eq:Tvec1b} gives
\begin{align}
\begin{pmatrix} T-1\\R \end{pmatrix}\!(E)
&=\frac{-\ii \mu}{p^3}\int \dd\xi\,\dd\eta
\sum_{n=1}^\infty \left(\Sigma_{p_+}^{\,n-1}\right)\!(\xi,\eta)\nonumber\\
&\qquad\times V(\eta)
\begin{bmatrix}
    E\\[2pt]
    m\,\ee^{2\ii p^3\eta}
\end{bmatrix}.
\label{eq:TR_explicit}
\end{align}

\begin{figure}[t]
  \centering
  \includegraphics[width=0.95\columnwidth]{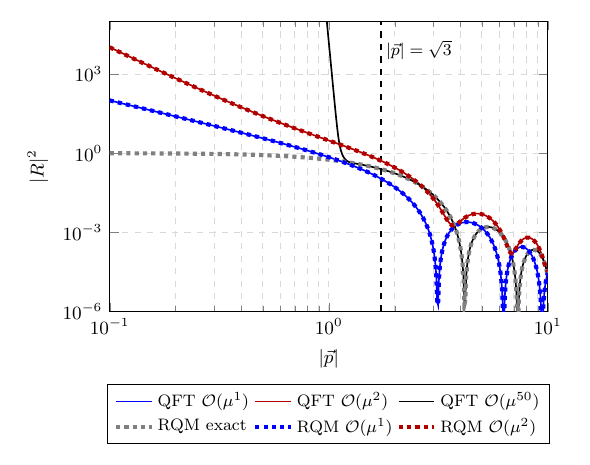}
  \caption{Reflection probability $|R|^2$ as function of the longitudinal amplitude $p^3\equiv|\vec p|$ for a particle of mass $m=1$ through a rectangular
    barrier with width $L=1$ and height $V_0=1$. The vertical dashed line corresponds to the momentum barrier height $|\vec{p}|=\sqrt{3}$ in momentum. Solid lines show partial sums of the perturbative expansion \eqref{eq:TR_explicit} up to order $\mu^N$ for $N=1,2$ and 50.
    The gray dotted line shows the exact result from RQM [following from Eq.~\eqref{eq:R}], while the blue and red dotted lines are the RQM reflection at order $\mathcal{O}(\mu^1)$ and $\mathcal{O}(\mu^2)$, respectively.
}
  \label{fig:convergence-plots}
\end{figure}

Figure~\ref{fig:convergence-plots} shows the reflection probability $|R|^2$ for a square barrier at order $N$ in the coupling constant $\mu$. 
Predictions from Eq.~\eqref{eq:TR_explicit} match the RQM results at all order above the barrier, but do not converge toward the exact RQM transmission below the barrier, as shown by the divergence for $N=50$.  
% }
The  divergence of the series in the
tunnelling regime, confirms that a non-perturbative treatment is
required, consistent with general arguments on the divergence of
perturbation theory in quantum field
theory~\cite{dyson_divergence_1952,lipatov_divergence_1977,suslov_divergent_2005}.
 
%    Above-barrier ($|p^3|>\sqrt{3}$): convergence to the exact RQM result    (dashed).    Below-barrier ($|p^3|<\sqrt{3}$): divergence, necessitating resummation.}

\subsection{Schwinger--Dyson equation for the densities}
\label{sec:SDE}

% {\color{blue}
Equation~\eqref{eq:TR_explicit} can be written as $(T-1,R)^T=\int \dd\xi \,\vec{\tau}(\xi)$ with 
\begin{equation}
    \vec{\tau}(\xi)=(\tau_T(\xi),\tau_R(\xi))^T\equiv \int\dd\eta\,\sum_{n=1}^{\infty}(\Sigma_{p_+}^{n-1})(\xi,\eta)\,\vec{\tau}_1(\eta)\label{eq:neumann}
\end{equation}
and 
\begin{equation}
    \vec{\tau}_1(\eta)=\frac{-\ii\mu}{p^3}V(\eta)
    \begin{pmatrix}E\\ m\,\ee^{2\ii p^3\eta}\end{pmatrix}.
    \label{eq:tau1}
\end{equation}
 The Neumann series~\eqref{eq:neumann}, though defined order by order,
sums to a function satisfying the self-consistent equation
%\begin{equation}
%  \vec{\mathcal{T}}_\infty(\xi)
%  = \vec{\mathcal{T}}^{(1)}(\xi)
%  + \int\!\dd\eta\;\Sigma(\xi,\eta)\,\vec{\mathcal{T}}_\infty(\eta)\,,
%  \label{eq:SDE}
%\end{equation}
\begin{equation}
  \vec{\tau}(\xi)
  = \vec{\tau}_{1}(\xi)
  + \int\!\dd\eta\;\Sigma_{p_+}(\xi,\eta)\,\vec{\tau}(\eta)\,,
  \label{eq:SDE}
\end{equation}
% }
which is exact and non-perturbative.
Equation~\eqref{eq:SDE} is the position-space density analogue of the
Schwinger--Dyson equation that dresses the bare fermion
propagator~\cite{dyson_radiation_1949,schwinger_gauge_1951,bashir_dynamical_2012,curtis_truncating_1990}.
Whereas the propagator equation is a self-referential relation on a
four-dimensional matrix-valued function and is in general intractable,
its one-dimensional density analogue~\eqref{eq:SDE} closes on a
two-component vector and, as we now show, is exactly solvable.

\subsection{Reduction to an ODE system}
\label{sec:ODE}
 
% Introduce auxiliary antiderivatives
% \begin{equation}
%   \Fcal_T(\xi) = \int_{-\infty}^\xi\!\dd\eta\;T_\infty(\eta)\,,
%   \qquad
%   \Fcal_R(\xi) = \int_\xi^\infty\!\dd\eta\;R_\infty(\eta)\,,
%   \label{eq:aux}
% \end{equation}
% satisfying $\Fcal_T'=T_\infty$, $\Fcal_R'=-R_\infty$,
% with boundary conditions $\Fcal_T(-\infty)=0$ and $\Fcal_R(+\infty)=0$.
% Expanding~\eqref{eq:SDE} with the explicit density~\eqref{eq:Sigma} and
% differentiating once with respect to $\xi$ converts the integral equation
% into the coupled first-order ODE system

Introduce auxiliary antiderivatives
% {\color{blue}
\begin{equation}
  \Fcal_T(\xi) = \int_{-\infty}^\xi\!\dd\eta\;\tau_T(\eta)\,,
  \qquad
  \Fcal_R(\xi) = \int_\xi^\infty\!\dd\eta\;\tau_R(\eta)\,,
  \label{eq:aux}
\end{equation}
satisfying $\Fcal_T'=\tau_T$, $\Fcal_R'=-\tau_R$,
% }
with boundary conditions $\Fcal_T(-\infty)=0$ and $\Fcal_R(+\infty)=0$.
The step functions in the kernel~\eqref{eq:Sigma} render the $\eta$-integral
in~\eqref{eq:SDE} a pair of antiderivatives of exactly this form, as shown in Appendix \ref{app:ODE}, so that
differentiating once with respect to $\xi$ collapses the integral equation
into the coupled first-order ODE system

\begin{widetext}
\begin{equation}
  \frac{\dd}{\dd\xi}\vec\Fcal
  + \frac{\ii\mu\, V(\xi)}{p^3}
    \begin{bmatrix}
      E & m\,\ee^{-2\ii p^3\xi}\\[4pt]
      -m\,\ee^{2\ii p^3\xi} & -E
    \end{bmatrix}
    \vec\Fcal
  + \frac{\ii\mu\, V(\xi)}{p^3}
    \begin{bmatrix}
      E\\[4pt]-m\,\ee^{2\ii p^3\xi}
    \end{bmatrix}
  = 0\,,
  \label{eq:ode}
\end{equation}
\end{widetext}
 
\noindent
where $\vec\Fcal=(\Fcal_T,\Fcal_R)^T$.
The physical amplitudes are recovered from the boundary values:
\begin{equation}
  T = 1 + \Fcal_T(+\infty)\,,
  \qquad
  R = \Fcal_R(-\infty)\,.
  \label{eq:ode_amplitudes}
\end{equation}
Equation~\eqref{eq:ode} is the central result of this work.
It constitutes a well-posed two-point boundary value problem, solvable by
standard Runge--Kutta
shooting~\cite{weideman_split-step_1986} for any
$V(\xi)$ and admitting exact closed-form solution for potentials with
piecewise-constant coefficients.

%=============================================================
\section{Application to rectangular barrier}
\label{sec:results}
%=============================================================
 
For the rectangular barrier $V(z)=V_0\,\mathbf{1}_{[0,L]}(z)$, the ODE
system~\eqref{eq:ode} has constant coefficients inside the barrier.
Solving by standard matrix-exponential methods with matching conditions at
$z=0$ and $z=L$ (Appendix~\ref{app:rect}) yields the closed-form amplitudes
\begin{align}
  T &= \frac{\kappa\,\ee^{-\ii p^3 L}}
           {\kappa\cos\kappa L
            - \ii\!\left(p^3 - \dfrac{V_0 E}{p^3}\right)\!\sin\kappa L}\,,
  \label{eq:T}\\[6pt]
  R &= \frac{-\ii\dfrac{V_0 m}{p^3}\sin\kappa L}
           {\kappa\cos\kappa L
            - \ii\!\left(p^3-\dfrac{V_0 E}{p^3}\right)\!\sin\kappa L}\,,
  \label{eq:R}
\end{align}
where $\kappa=\sqrt{(E-V_0)^2-m^2}$.
% In the tunnelling regime $(p^0-V_0)^2<m^2$, $\kappa$ becomes purely imaginary; writing $\kappa = \ii|\kappa|$ gives $\cos\kappa L = \cosh|\kappa|L$ and $\sin\kappa L = \ii\sinh|\kappa|L$, so the amplitudes are exponentially suppressed as expected.
In the tunnelling regime $(E-V_0)^2<m^2$, $\kappa$ becomes purely
imaginary; writing $\kappa=\ii|\kappa|$ gives $\cos(\kappa L)=\cosh(|\kappa|L)$
and $\sin(\kappa L)=\ii\sinh(|\kappa|L)$, whereupon the numerator of~\eqref{eq:T}
remains $O(1)$ while the denominator grows as $\cosh|\kappa|L$, so
$|T|\sim\ee^{-|\kappa|L}$ is exponentially suppressed as expected.

Equations~\eqref{eq:T}--\eqref{eq:R} are in exact agreement with the
standard RQM results obtained by matching Dirac spinor wavefunctions at the
barrier
boundaries~\cite{de_leo_above_2006,calogeracos_history_1999,cotaescu_applying_2007},
and satisfy unitarity $|T|^2+|R|^2=1$ identically.
%The agreement is verified numerically by the Runge--Kutta solution
%of~\eqref{eq:ode} across the full parameter space.
% The agreement is verified numerically by the Runge--Kutta solution
% of~\eqref{eq:ode} across the full parameter space.
The agreement is confirmed by the Runge--Kutta solution of~\eqref{eq:ode}
over a broad range of $p^3$ and $V_0$, both above and below the barrier.

%=============================================================
\section{Discussion}
\label{sec:conclusions}
%=============================================================
 
We have derived a non-perturbative QFT framework for spinor tunnelling by
resumming the all-order Feynman diagram expansion into a
Schwinger--Dyson-type equation on position-space densities and have further
reduced this to the ODE system~\eqref{eq:ode}.
The framework (i)~handles arbitrary smooth potentials numerically,
(ii)~yields exact amplitudes for piecewise-constant potentials,
and (iii)~reproduces RQM results in the single-particle sector.

The $2\times 2$ spinor matrix structure, in the recursion~\eqref{eq:recursion}
and the ODE~\eqref{eq:ode}, couples the transmission and reflection
channels at each step.
This coupling is a distinguishing feature of the spinor case: the
scalar resummation of Ref.~\cite{zielinski_tunnelling_2024} yields
a single integral equation, whereas the matrix-valued Dirac propagator
numerator $\slashed{q}+m$ generates the off-diagonal terms that mix
forward- and backward-propagating modes.
% Physically, these off-diagonal entries encode
% the spin-flip processes that are absent for scalar particles and that
% underlie the Klein paradox phenomenology~\cite{klein_reflexion_1929,
% holstein_kleins_1998,dombey_seventy_1999}.
Physically, these off-diagonal entries couple the forward- and
backward-propagating solutions of the Dirac equation. 
The resulting mixing is generated by the spinor components and has no analogue in the scalar case.

The primary motivation for a QFT treatment is access to corrections beyond
RQM.
These include vacuum
polarisation~\cite{uehling_polarization_1935,schwinger_gauge_1951},
vertex dressing~\cite{schwinger_quantum-electrodynamics_1948,
bonciani_vertex_2003} and the pair-production
effects that resolve the Klein
paradox~\cite{hansen_kleins_1981}.
Radiative corrections have been shown to increase tunnelling probabilities
for charged particles through barrier
renormalisation~\cite{flambaum_radiation_1999} and 
a formalism to compute analogous corrections for scalar
fields has recently been proposed~\cite{zielinski_quantum_2024}; the present
work provides the spinor foundation for this program.

A complete calculation of tunnelling incorporating a dynamical photon field
in~\eqref{eq:lagrangian} and one-loop (and higher) QED radiative corrections
to beyond-RQM tunnelling is the natural next step.
The ODE formulation~\eqref{eq:ode} may be better suited to facilitate this extension. For instance, dressed propagator terms or vertex corrections could be introduced to modify the ODE coefficients without altering its overall structure.
Such corrections are relevant not only to fundamental tests of
QED~\cite{karshenboim_precision_2005,lindgren_many-body_2006}, but also to tunnelling phenomena in nuclear
physics~\cite{simenel_nuclear_2025} and strong-field
QED~\cite{fedotov_advances_2023,hattori_strong-field_2023}.

\begin{acknowledgments}
We are grateful to A.G. Williams for valuable discussions and insights. 
M.F. acknowledges the support of the Australian National University through the Dunbar Physics
Honours Scholarship. This research was supported by the Australian Government through the Australian Research Council Centre of Excellence for Dark Matter Particle
Physics (CE200100008).
\end{acknowledgments}

%=============================================================
\appendix
\section{Derivation of the recursion relation and amplitude seed}
\label{app:derivation}
%=============================================================

\subsection{Scattering densities}
\label{app:densities}
\begin{widetext}
 
Inserting the factorised potential~\eqref{eq:Vtilde} at each vertex of
Eq.~\eqref{eq:nth_amplitude} gives the explicit form of the scattering
density defined in Eq.~\eqref{eq:density_def},
\begin{equation}
\mathcal{S}_{kp}^{(n)}(\vec{z})=(-\ii\mu)^n
     \int\!\frac{\dd^4 q_1}{(2\pi)^4}\cdots\frac{\dd^4q_{n-1}}{(2\pi)^4}\,\,
    \bar u(k)\,\gamma^0\,
    \left(\prod_{i=0}^{2}2\pi\delta(k^i{-}q_{n-1}^i)\right)\,V(z_n)\,
    \ee^{-\ii(k^3-q_{n-1}^3)\,z_n}\,\,
    P^{(n-1)}_{\vec{q},\vec{z}} \,u(p)\,,
    \label{eq:density_explicit}
\end{equation}
where
\begin{equation}
P^{(n-1)}_{\vec{q},\vec{z}}\equiv\mathcal{P}\left[\prod_{j=1}^{n-1}G_F(q_j)\,\gamma^0\,
    \left(\prod_{i=0}^{2}2\pi\delta(q_j^i-q^i_{j-1})\right)
    V(z_j)\,\ee^{-\ii (q_j^3-q_{j-1}^3) z_{j}}\right]\,.
    \label{eq:Pdef}
\end{equation}
Isolating the outermost internal momentum $q_{n-1}$,
\begin{equation}
\mathcal{S}_{kp}^{(n)}(\vec{z})
  = (-\ii\mu)^n
    \int\!\frac{\dd^4 q_1}{(2\pi)^4}\cdots\frac{\dd^4q_{n-2}}{(2\pi)^4}\,\,
    \bar u(k)\,\Ical(z_n,z_{n-1},k,q_{n-2})\,
 P^{(n-2)}_{\vec{q},\vec{z}} \, u(p)\,,
  \label{eq:density_I}
\end{equation}
where
\begin{align}
  \Ical(z_n,z_{n-1},k,q_{n-2})
  &=  \int\!\frac{\dd^4 q_{n-1}}{(2\pi)^4}\left(\prod_{i=0}^{2}(2\pi)^2\delta(k^i{-}q_{n-1}^i)\delta(q^i_{n-1}{-}q_{n-2}^i)\right)
    V(z_n)\,V(z_{n-1})\;
    \ee^{\ii q_{n-2}^3 z_{n-1} - \ii k^3 z_n}\nonumber\\
&\,\,\,\,\,\times\;    \gamma^0\,\Gcal_F(q_{n-1})\,\ee^{\ii q^3_{n-1}(z_n-z_{n-1})}\,\gamma^0\nonumber\\
&= \left(\prod_{i=0}^{2}2\pi\delta(k^i{-}q_{n-2}^i)\right)\;
    V(z_n)\,V(z_{n-1})\;
    \ee^{\ii q_{n-2}^3 z_{n-1} - \ii k^3 z_n}\;
    \gamma^0\,\hat\Gcal(\hat{k},z_n-z_{n-1})\,\gamma^0\,,
  \label{eq:Idensity}
\end{align}
with $\hat\Gcal$ the longitudinal Fourier transform of the propagator
defined in Eq.~\eqref{eq:Ghat_def}.
\end{widetext}
 
\subsection{Recursion relation}
\label{app:recursion}
\begin{widetext}
Sandwiching Eq.~\eqref{eq:Idensity} against an on-shell spinor and using
Eq.~\eqref{eq:G_result_simp} gives
\begin{align}
  \bar u(k_\pm)\,\Ical(z_n,z_{n-1},k_\pm,q_{n-2})
  &= \frac{V(z_n)}{k^3}\,\left(\prod_{i=0}^{2}2\pi\delta(k^i{-}q_{n-2}^i)\right)\,
     V(z_{n-1})\left[
     \Theta(\pm z_n{\mp}z_{n-1})\,k^0\,
     \ee^{\ii(q_{n-2}^3\mp k^3)\,z_{n-1}}\,\bar u(k_\pm)
     \;\right.
  \nonumber\\
  &\quad\left.+\;\ee^{\mp2\ii k^3 z_n}\,
     \Theta(\mp z_{n}{\pm}z_{n-1})\,m\,
     \ee^{\ii(q_{n-2}^3\pm k^3)\,z_{n-1}}\,\bar u(k_\mp)
  \right]\,
     \gamma^0\,.
  \label{eq:Idensity_expanded}
\end{align}
Substituting Eq.~\eqref{eq:Idensity_expanded} into
Eq.~\eqref{eq:density_I} and comparing with Eq.~\eqref{eq:density_explicit}
taken at order $n-1$, we get
\begin{equation}
\mathcal{S}_{k_\pm p}^{(n)}(\vec{z})
=-\ii\mu\frac{V(z_n)}{k^3}\left[\Theta(\pm z_n\mp z_{n-1})\,k^0\,\mathcal{S}_{k_\pm p}^{(n-1)}(\vec{z})+\Theta(\mp z_{n}\pm z_{n-1})\,m\,\ee^{\mp2\ii k^3z_n}\,\mathcal{S}_{k_\mp p}^{(n-1)}(\vec{z})\right]\,,
\label{eq:Spm}
\end{equation}
which is the recursion relation~\eqref{eq:recursion} in
matrix form.
\end{widetext}

\subsection{First-order seed and final-state integration}
\label{app:seed}
\begin{widetext}

The iteration~\eqref{eq:TnT1} terminates on the first-order seed
\begin{equation}
\Tvec_{kp}^{(1)}(\xi)=
  \begin{bmatrix}
    \mathcal{S}^{(1)}_{k_+ p}(\xi)\\[4pt]
    \mathcal{S}^{(1)}_{k_- p}(\xi)
  \end{bmatrix}\,,
  \label{eq:Tvec1}
\end{equation}
which is determined from the first-order S-matrix element,
\begin{equation}
     \langle k|S|p\rangle^{(1)}= \int \dd z_1 \,\mathcal{S}_{kp}^{(1)}(z_1)=
           \begin{tikzpicture}[baseline={([yshift=-.5ex]current bounding box.center)},
                          scale=0.8, transform shape]
        \begin{feynman}
          \vertex (v);
          \vertex[above=1.4cm of v, inner sep=0pt, outer sep=0pt] (x)
                 {\Large$\otimes$};
          \vertex[left=1.4cm of v] (a);
          \vertex[right=1.4cm of v] (b);
          \diagram*{
            (a)--[fermion, edge label=$p$](v)
                --[fermion, edge label=$k$](b),
            (v)--[photon](x)
          };
        \end{feynman}
      \end{tikzpicture}
=\; -\ii\mu\,\tilde V(k-p)\,\bar{u}(k)\gamma^0u(p)\,.      \label{eq:Smat1}
\end{equation}
Using Eq.~\eqref{eq:Vtilde}, we get
\begin{equation}
        \mathcal{S}_{kp}^{(1)}(\xi)= -\ii\mu\left(\prod_{i=0}^2 2\pi\delta(k^i-p^i)\right)
\, V(\xi)\,\ee^{-\ii(k^3-p^3)\xi}\,\bar{u}(k)\gamma^0u(p)\,.
\label{eq:S1density}
\end{equation}
Substituting in Eq.~\eqref{eq:Tvec1} gives
\begin{align}
\Tvec_{kp}^{(1)}(\xi)&=     -\ii\mu\left(\prod_{i=0}^2 2\pi\delta(k^i-p^i)\right)
\, V(\xi) \begin{bmatrix}
    \ee^{\ii(p^3-k^3)\xi}\,\bar{u}(k_+)\\[4pt]
    \ee^{\ii(p^3+k^3)\xi}\,\bar{u}(k_-)
  \end{bmatrix}\,\gamma^0u(p)\nonumber\\
  &=\frac{-2\ii\mu E}{p^3}(2\pi)^3\delta(k^1)\delta(k^2)V(\xi)\left(\delta(k^3-p^3)
  \begin{bmatrix}
   E\\[4pt]
    \ee^{2\ii p^3\xi}\,m
  \end{bmatrix}
  +\delta(k^3+p^3)
  \begin{bmatrix}
       \ee^{2\ii p^3\xi}\,m\\[4pt]
    E
  \end{bmatrix}\right)\,,
  \label{eq:Tvec1b}
\end{align}
where $\delta(k^0-p^0)=\frac{p^0}{p^3}[\delta(k^3-p^3)+\delta(k^3+p^3)]$ (valid on the support of $\delta(k^1)\delta(k^2)$), Eqs.~\eqref{eq:spinor_ips}, and $p=(E,0,0,p^3)$ have been used in the last step.
 The transmission and reflection amplitudes~\eqref{eq:TR} then read
\begin{align}
\begin{pmatrix} T\\R
\end{pmatrix}(E)&=\int\frac{\dd^3k}{(2\pi)^3}\frac{1}{2k^0}
\left(\langle k|S|p\rangle^{(0)}+\sum_{n=1}^\infty\int\dd z_1\cdots\dd z_n\, \mathcal{S}_{kp}^{(n)}(\vec{z})\right)\begin{bmatrix}
\Theta(k^3)\\
\Theta(-k^3)
\end{bmatrix}\,,
\label{eq:TRassembly}
\end{align}
where the zeroth order S-matrix element is simply,
\begin{equation}
  \langle k|S|p\rangle^{(0)}=2E(2\pi)^3\delta^{(3)}(\mathbf{k}-\mathbf{p})\, ,
  \label{eq:zeroth_order}
\end{equation}
which carries no power of $V$ and furnishes only the unit contribution to
$T$.
% {\color{blue}
Using $\mathcal{S}_{kp}^{(n)}=\mathcal{S}_{k_+p}^{(n)}$ and $\int \dd k^3\,\mathcal{S}_{kp}^{(n)}\Theta(-k^3) = \int \dd k^3\,\mathcal{S}_{k_-p}^{(n)}\Theta(k^3)$, we can factorise $\Theta(k^3)$, giving
\begin{align}
\begin{pmatrix} T-1\\R
\end{pmatrix}(E)&=\int\frac{\dd^3k}{(2\pi)^3}\frac{\Theta(k^3)}{2k^0}
\sum_{n=1}^\infty\int\dd z_1\cdots\dd z_n\, \begin{bmatrix}
\mathcal{S}_{k_+p}^{(n)}(\vec{z})\\
\mathcal{S}_{k_-p}^{(n)}(\vec{z})
\end{bmatrix}=\int\frac{\dd^3k}{(2\pi)^3}\frac{\Theta(k^3)}{2k^0}
\sum_{n=1}^\infty\int \dd\xi\,\left(\Sigma_k^{\,n-1}\cdot\Tvec_{kp}^{(1)}\right)(\xi)\,,\nonumber
%\label{eq:TRassembly}
\end{align}
where Eqs.~\eqref{eq:Tvec_def} and \eqref{eq:TnT1} have been used in the last step to give Eq.~\eqref{eq:TRassembly_main}.

\section{Reduction to an ODE}
\label{app:ODE}
%=============================================================

From Eq.~\eqref{eq:Sigma}, we can write 
\begin{align}
 \int\!\dd\eta\;\Sigma_{p_+}(\xi,\eta)\,\vec{\tau}(\eta)&=
   \frac{-\ii\mu\, V(\xi)}{p^3}\, \int_{-\infty}^\infty\!\dd\eta\;
  \begin{bmatrix}
    E\,\Theta(\xi{-}\eta) &
    m\,\ee^{-2\ii p^3\xi}\,\Theta(\eta{-}\xi)\\[4pt]
    m\,\ee^{2\ii p^3\xi}\,\Theta(\xi{-}\eta) &
    E\,\Theta(\eta{-}\xi)
  \end{bmatrix}\,\vec{\tau}(\eta)\\
  &=
   \frac{-\ii\mu\, V(\xi)}{p^3}\, 
  \begin{bmatrix}
    E \int_{-\infty}^\xi\!\dd\eta &
    m\,\ee^{-2\ii p^3\xi}\int_{\xi}^\infty\!\dd\eta\\[4pt]
    m\,\ee^{2\ii p^3\xi}\int_{-\infty}^\xi\!\dd\eta &
    E\int_{\xi}^\infty\!\dd\eta
  \end{bmatrix}\,\vec{\tau}(\eta)\\
    &=
   \frac{-\ii\mu\, V(\xi)}{p^3}\, 
  \begin{bmatrix}
    E &
    m\,\ee^{-2\ii p^3\xi}\\[4pt]
    m\,\ee^{2\ii p^3\xi} &
    E
  \end{bmatrix}\,
  \begin{pmatrix}
      \int_{-\infty}^\xi\!\dd\eta \,\tau_T(\eta)\\\int_\xi^\infty\!\dd\eta \,\tau_R(\eta)
  \end{pmatrix}\,.
    \label{eq:ODE1}
\end{align}
% }
\end{widetext}
% {\color{blue}
We see from the definitions of the auxiliary antiderivatives in Eq.~\eqref{eq:aux} that the vector in the r.h.s. of Eq.~\eqref{eq:ODE1} is simply $\vec\Fcal=(\Fcal_T,\Fcal_R)^T$. 
Using $\vec{\tau}(\xi)=\frac{\dd}{\dd\xi}\left(\Fcal_T,-\Fcal_R\right)^T$, Eq.~\eqref{eq:SDE} can be expressed as 
\begin{align}
\begin{pmatrix}
    \Fcal_T'(\xi)\\ -\Fcal_R'(\xi)
\end{pmatrix}    &=
\vec{\tau}_1  - \frac{\ii\mu V(\xi)}{p^3}
  \begin{bmatrix}
    E &
    m\ee^{-2\ii p^3\xi}\\[4pt]
    m\ee^{2\ii p^3\xi} &
    E
  \end{bmatrix}\,
\begin{pmatrix}
    \Fcal_T(\xi)\\ \Fcal_R(\xi)
\end{pmatrix}    .
    \label{eq:ODE2}
\end{align}
Using the explicit expression for $\vec\tau_1$ in Eq.~\eqref{eq:tau1} and rearranging gives Eq.~\eqref{eq:ode}.
% }

%=============================================================
\section{Exact solution for the rectangular barrier}
\label{app:rect}
%=============================================================

Inside the barrier $\xi\in[0,L]$ the potential is constant, $V(\xi)=V_0$,
and the ODE system~\eqref{eq:ode} has constant coefficients.
Outside the barrier the potential vanishes and~\eqref{eq:ode} gives
$\dd\vec{\Fcal}/\dd\xi=0$, so $\vec{\Fcal}$ is piecewise constant
for $\xi<0$ and $\xi>L$.
The physical boundary conditions are
\begin{equation}
  \Fcal_T(\xi\to-\infty)=0\,,\qquad
  \Fcal_R(\xi\to+\infty)=0\,,
  \label{eq:bc_phys}
\end{equation}
so that $\Fcal_T(\xi)=0$ for all $\xi<0$ and
$\Fcal_R(\xi)=0$ for all $\xi>L$.
Together with continuity at $\xi=0$ and $\xi=L$ this fixes the
solution uniquely.

\paragraph{Autonomous form.}
The oscillatory coupling in~\eqref{eq:ode} is removed by the substitution
$\tilde\Fcal_R(\xi)=\ee^{-2\ii p^3\xi}\,\Fcal_R(\xi)$,
giving the autonomous system
\begin{equation}
  \frac{\dd}{\dd\xi}
  \begin{bmatrix}\Fcal_T\\\tilde\Fcal_R\end{bmatrix}
  = \ii\,\mathbf{M}
  \begin{bmatrix}\Fcal_T\\\tilde\Fcal_R\end{bmatrix}
  + \ii\,\vec{s}\,,
  \label{eq:ode_const}
\end{equation}
where (setting $\mu=1$, absorbed into $V_0$)
\begin{equation}
  \mathbf{M}
  = \begin{bmatrix}
      -\dfrac{V_0 E}{p^3} & -\dfrac{V_0 m}{p^3} \\[6pt]
      \dfrac{V_0 m}{p^3} & \dfrac{V_0 E}{p^3} - 2p^3
    \end{bmatrix},
  \qquad
  \vec{s}
  = \begin{bmatrix}-\dfrac{V_0 E}{p^3}\\[6pt] \dfrac{V_0 m}{p^3}\end{bmatrix}.
  \label{eq:M_and_s}
\end{equation}
The second diagonal entry of $\mathbf{M}$ contains a kinematic shift
$-2p^3$ generated by the substitution that is independent of the
coupling and cannot be absorbed into the overall $V_0/p^3$ prefactor.

\paragraph{Eigendecomposition.}
Using the on-shell relation $(p^3)^2=E^2-m^2$, one finds
$\mathrm{tr}\,\mathbf{M}=-2p^3$ and
$\det\mathbf{M}=(p^3)^2-\kappa^2$, so the eigenvalues of $\mathbf{M}$ are
\begin{equation}
  \lambda_\pm = -p^3 \pm \kappa\,,\qquad
  \kappa \equiv \sqrt{(E-V_0)^2-m^2}\,,
  \label{eq:eigenvalues}
\end{equation}
with corresponding eigenvectors
\begin{equation}
  \vec{v}_\pm
  = \begin{bmatrix}
      \dfrac{V_0 m}{p^3} \\[6pt]
      p^3 - \dfrac{V_0 E}{p^3} \mp \kappa
    \end{bmatrix}.
  \label{eq:eigenvectors}
\end{equation}
A particular (constant) solution
$\vec{\Fcal}_{\mathrm{p}}=-\mathbf{M}^{-1}\vec{s}$
of~\eqref{eq:ode_const} is obtained by direct inversion, which on use of
the on-shell relation reduces to the simple form
\begin{equation}
  \vec{\Fcal}_{\mathrm{p}}
  = \begin{bmatrix}-1\\[2pt] 0\end{bmatrix}.
  \label{eq:particular}
\end{equation}
The general solution of~\eqref{eq:ode_const} for $\xi\in[0,L]$ is therefore
\begin{equation}
  \begin{bmatrix}\Fcal_T(\xi)\\\tilde\Fcal_R(\xi)\end{bmatrix}
  = \vec{\Fcal}_{\mathrm{p}}
  + c_+\,\vec{v}_+\,\ee^{\ii\lambda_+\xi}
  + c_-\,\vec{v}_-\,\ee^{\ii\lambda_-\xi}\,.
  \label{eq:gen_sol}
\end{equation}

\paragraph{Boundary matching.}
The boundary condition $\Fcal_T(\xi\to-\infty)=0$ together with
continuity at $\xi=0$ gives $\Fcal_T(0^-)=0$. The boundary
condition $\Fcal_R(\xi\to+\infty)=0$ together with continuity at
$\xi=L$ gives $\tilde\Fcal_R(L^+)=0$.
Defining
\begin{equation}
  \alpha_\pm \equiv p^3 - \frac{V_0 E}{p^3} \mp \kappa\,,
  \qquad
  \zeta_\pm \equiv \ee^{\ii\lambda_\pm L}
  = \ee^{-\ii p^3 L}\,\ee^{\pm\ii\kappa L}\,,
\end{equation}
the two matching conditions read
\begin{align}
  -1 + \frac{V_0 m}{p^3}(c_+ + c_-) &= 0\,,
  \label{eq:match_T}\\
  c_+\,\alpha_+\,\zeta_+ + c_-\,\alpha_-\,\zeta_- &= 0\,.
  \label{eq:match_R}
\end{align}
Equation~\eqref{eq:match_T} gives $c_++c_-=p^3/(V_0 m)$ and combined
with~\eqref{eq:match_R} fixes $c_\pm$ uniquely.

\paragraph{Solving for $T$ and $R$.}
Substituting into~\eqref{eq:gen_sol} and using the identity
\begin{equation}
  \alpha_+\alpha_- 
  = \left(p^3 - \frac{V_0 E}{p^3}\right)^{\!2} - \kappa^2
  = \frac{V_0^2 m^2}{(p^3)^2}\,,
\end{equation}
which follows from the on-shell relation, yields after standard
manipulation
\begin{align}
  \Fcal_T(L^+) &= -1 + \frac{\kappa\,\ee^{-\ii p^3 L}}
       {\kappa\cos\kappa L
        -\ii\!\left(p^3-\dfrac{V_0 E}{p^3}\right)\!\sin\kappa L}\,,\\[4pt]
  \Fcal_R(0^-) &= \frac{-\ii\dfrac{V_0 m}{p^3}\sin\kappa L}
       {\kappa\cos\kappa L
        -\ii\!\left(p^3-\dfrac{V_0 E}{p^3}\right)\!\sin\kappa L}\,.
\end{align}
The physical amplitudes follow directly from~\eqref{eq:ode_amplitudes}:
\begin{align}
  T &= 1 + \Fcal_T(+\infty) 
     = \frac{\kappa\,\ee^{-\ii p^3 L}}
       {\kappa\cos\kappa L
        -\ii\!\left(p^3-\dfrac{V_0 E}{p^3}\right)\!\sin\kappa L}\,,
  \label{eq:T_app}\\[4pt]
  R &= \Fcal_R(-\infty)
     = \frac{-\ii\dfrac{V_0 m}{p^3}\sin\kappa L}
       {\kappa\cos\kappa L
        -\ii\!\left(p^3-\dfrac{V_0 E}{p^3}\right)\!\sin\kappa L}\,,
  \label{eq:R_app}
\end{align}
where the equalities $\Fcal_T(+\infty)=\Fcal_T(L^+)$ and
$\Fcal_R(-\infty)=\Fcal_R(0^-)$ hold because $\vec{\Fcal}$ is constant
outside $[0,L]$.

\bibliography{refs}% Produces the bibliography via BibTeX.

\end{document}